Astro2020 APC White Paper

# Findings and Recommendations from the American Astronomical Society (AAS) Committee on the Status of Women in Astronomy: Advancing the Career Development of Women in Astronomy

Thematic Area: State of the Profession Considerations

Principal Author:

Nicolle Zellner, Albion College, nzellner@albion.edu

Co-Authors:

JoEllen McBride, University of Pennsylvania, joellen.mcbride@gmail.com
Nancy Morrison, University of Toledo, nancyastro126@gmail.com
Alice Olmstead, Texas State University, alice.olmstead@txstate.edu
Maria Patterson, High Alpha, maria.t.patterson@gmail.com
Gregory Rudnick, University of Kansas, grudnick@ku.edu
Aparna Venkatesan, University of San Francisco, avenkatesan@usfca.edu
Heather Flewelling, University of Hawaii Institute for Astronomy, heather@ifa.hawaii.edu
David Grinspoon, Planetary Science Institute, grinspoon@psi.edu
Jessica D. Mink, Harvard University Center for Astrophysics, jmink@cfa.harvard.edu
Christina Richey, Jet Propulsion Laboratory, christina.r.richey@jpl.nasa.gov
Angela Speck, University of Missouri, speckan@missouri.edu
Cristina A. Thomas, Northern Arizona University, Cristina.Thomas@nau.edu
Sarah E. Tuttle, University of Washington, tuttlese@uw.edu

Endorsers:
Kim Coble, San Francisco State University, kcoble@sfsu.edu
Lia Corrales, University of Michigan, liac@umich.edu
Stephen Lawrence, Hofstra University, Stephen.Lawrence@hofstra.edu

*The views expressed in this white paper are not necessarily the views of the AAS, its Board of Trustees, or its membership.*

**Executive Summary**

The Committee on the Status of Women in Astronomy (CSWA) is calling on federal science funding agencies, in their role as the largest sources of funding for astronomy in the United States, to take actions that will advance career development and improve workplace conditions for women and minorities in astronomy. Funding agencies can and should lead the charge to produce excellent diversity and inclusion outcomes in our field by the 2030 Astrophysics Decadal Survey. Anecdotal and quantitative evidence, gathered both by the CSWA and other groups, shows that many structural barriers to success remain in our community. We acknowledge the success of programs like NSF-INCLUDES and NSF-ADVANCE and endorse their continued work. We also recommend further action to remove barriers to success for women and minority astronomers. Key recommendations are:

- Federal agencies should push academic institutions to reconsider their support systems for scientists by requiring a high standard of pay and benefits.
- Federal agencies should fund research and programs that explore and implement strategies for improving the experiences of scientists.
- Federal agencies should require Principal Investigators (PIs) to provide plans for creating inclusive work environments and mentoring young scientists.

1.1 Purpose

*This paper proposes actionable, evidence-supported policies that will provide essential workplace and career support for astronomers.*

This paper will focus on career development issues, while a sister paper focuses on ending harassment. We acknowledge that these areas are connected, and addressing each will have profound impacts on the other. The contents of Sections 1.1 and 1.2 and the first paragraph of Section 1.3 are repeated in both papers.

The CSWA was created in 1979 and was charged with making practical recommendations to the American Astronomical Society council on what can be done to improve the status of women in astronomy. The CSWA's scope has expanded to include all bodies that influence the work lives of astronomers, including research facilities, academic institutions, the federal government, and others[1]. The CSWA is submitting two reports on the pressing issues affecting women and minority astronomers, and recommending policy changes that fit the needs of our community in response to the decadal survey's call for white papers on the state of the astronomy profession.

## 1.2 Methodology

In Spring 2019, the CSWA conducted a survey to assess astronomers' perspectives on policies in four areas of concern: harassment and bullying, creating inclusive environments, professional development, and ethics. The survey included 53 Likert-scale questions that allowed astronomers to rate the effectiveness of policy actions that could be undertaken by relevant stakeholders, and 17 free response opportunities that they used to explain their answers. We received over 340 responses to the survey. No personally identifiable information was collected. Although we acknowledge the disadvantage of being unable to categorize our respondents' opinions by demographics, we believe the anonymous nature of the survey increased astronomers' willingness to take the survey and write candid free responses. Relevant expertise and data in the field of STEM equity also inform our recommendations.

## 1.3 Progress for Women in Astronomy is Slowing

*The number of women earning astronomy PhDs is increasing, but at a decreasing rate. Majority (white) and well-represented minority (Asian) women attain higher ranks than underrepresented minority (African American and Hispanic) women.*

According to the National Science Foundation (NSF), the number of women earning physics and astronomy doctoral degrees is increasing, but at a decreasing rate. The number of women who received doctoral degrees in physics and astronomy increased by 21.3% from 2008 to 2012, but by only 8.1% from 2013 to 2017[2]. Additionally, the American Institute of Physics (AIP) reports consistent decreases in the number of women earning astronomy bachelor's degrees over the past ten years[3].

While more and more women are becoming astronomers, they continue to be underrepresented in key astronomical workplaces. According to a study of NASA mission teams, from 2001 through 2016, the percentage of women involved in spacecraft science teams has remained flat, at 15.8%, even while the number of women in the field has continued to increase[4].

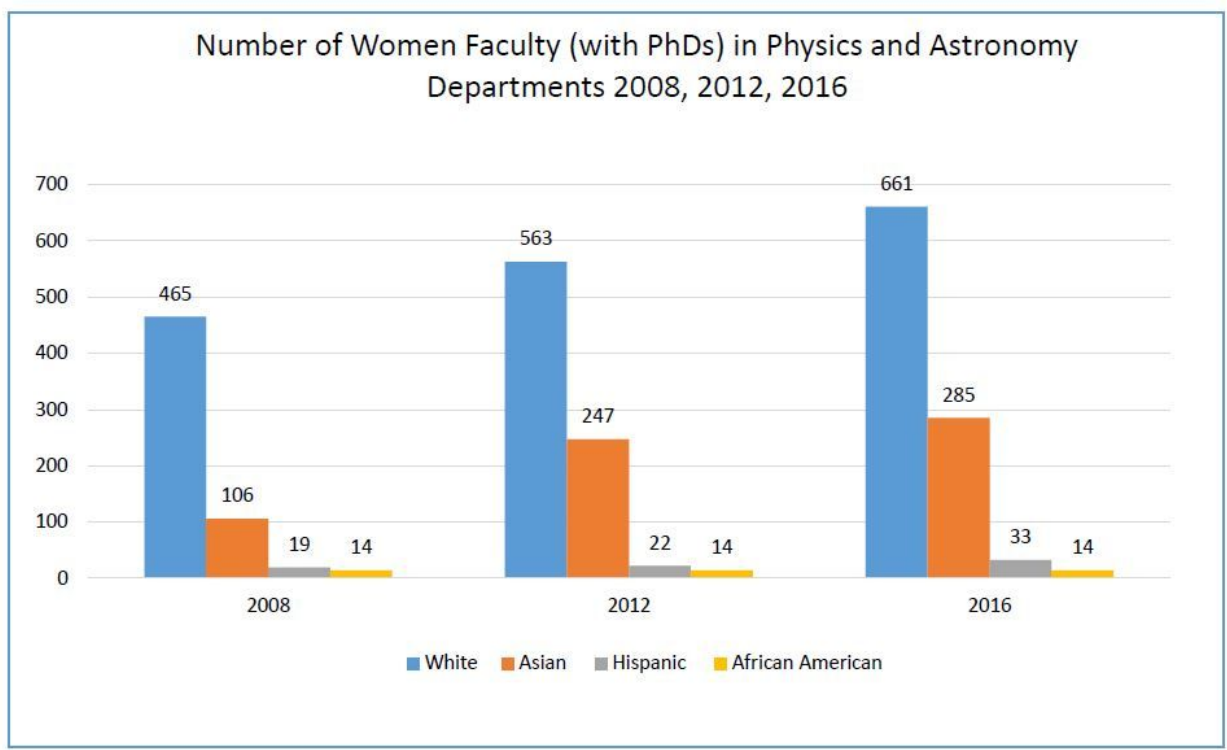

**Figure 1**. While the number of white women faculty increased by more than 40% over an 8-year period, the number of African American women faculty remained the same, a disturbing indicator of stagnation for this group. The overall percentage of women faculty remains around 20%[3].

The AIP also reports that while Hispanic women are entering astronomy in growing numbers, there has been no progress for African American women. Unequal progress for underrepresented minorities (URMs) is demonstrated by Figure 1, created using data from the AIP, which shows significant increases in the number of white and Asian women professors, but no increase in the number of African American women professors in physics and astronomy. The AIP included the following statement in their report: "In 2016, 5% of women who earned physics doctorates were Hispanic, and 3% were African-American. In that same year, 4% of women who earned an astronomy doctorate were Hispanic and 2% were African-American. Due to the small number of Hispanic and African-American women with doctorates, the NSF did not make data from previous years available to protect individuals from being identified."[3] In other words, there were so few Hispanic and African-American PhD recipients prior to 2016 that *they could all be identified by name.*

We also want to stress the importance of issues that are salient for LGBT scientists. Of respondents to the American Physical Society's 2016 LGBT Climate Survey, about 15% of LGBT men, 25% of LGBT women, and 30% of gender-nonconforming individuals characterized the overall climate of their department or division as "uncomfortable" or "very uncomfortable"[7]. A strong majority of respondents to our survey (74%) indicated that coordinating actions across advocacy groups to promote gender-neutral restrooms, lactation rooms, and other provisions for marginalized groups would be a "somewhat effective" or "very effective" strategy for creating inclusive environments.

Progress towards parity with men is slowing for women of all identity groups. Our survey responses demonstrate, and the literature shows, this is due to the inhospitable conditions in astronomy (and other STEM fields) that drive women, especially minority women, out of the field[5,6]. The urgency of the situation is summed up well by the words of one of our survey respondents, who wrote, "Ignoring intersectionality should not be treated as an option." Providing resources for those negatively impacted by the outsized effects of possessing one or more minority identity will improve the experiences of astronomers of all gender and racial identities, but will help those who are marginalized the most.

## 1.4 Initiate Policy Action to Advance Astronomers' Career Development
*Federal agencies can and should take action to create conditions where women can enjoy widespread success in astronomy.*

Our survey respondents expressed that the following factors caused them significant distress during their careers: lack of adequate pay for young scientists, the unavailability of resources such as healthcare benefits, and the defects of the inflexible structure of academia. Many pointed out that the dependence on one advisor for funding is not sustainable for scientists from lower income backgrounds. We think it is very possible that astronomy departments and academic institutions lack diversity because they do not address the needs of the diverse pool of students who have the ability and desire to persist with an astronomical career.

Additionally, diversity and inclusivity are vital because they enable the creation of diverse teams. Diverse teams work harder, provoke more thought, are more creative, and produce higher-quality scientific outcomes[8,9]. The presence of women on teams is associated with more diversified skill sets and higher social sensitivity, which leads to improved collaborative processes[10]. According to the National Academies of Science, Engineering, and Medicine's (NASEM) landmark report on harassment, there is less harassment in organizations led by diverse groups[11]. See also the two white papers "Providing a Timely Review of Input Demographics to Advisory Committees" and "Tying Research Funding to Progress in Inclusion" by Norman et al. in the Astro2020 APC white papers.

In addition to the CSWA, there are several other AAS committees working towards diversity and inclusion in astronomy. They are the Committee on the Status of Minorities in Astronomy (CSMA), the Committee on Sexual-Orientation and Gender Minorities in Astronomy (SGMA), the Working Group on Accessibility and Disability (WGAD), and the Professional Development and Professional Culture and Climate subcommittees within the Division of Planetary Sciences. They represent some of the areas of diversity beyond gender that are important to the growth of astronomy and all STEM fields.

H.R. 2528, The STEM Opportunities Act of 2019, states, "The Federal Government provides 55 percent of research funding at institutions of higher education and, through its grant-making policies, has had significant influence on institution[s] of higher education policies, including policies related to institutional culture and structure."[12] Federal agencies can and should use their influence to improve the structure of academic institutions and remove barriers to success for the astronomy workforce, leading to more and better science.

## 2. Career Support
*Federal agencies should take action to improve the management and funding structures that are meant to support scientists, to fund research that explores strategies for improving the experiences of scientists, and to provide more and better opportunities for mentorship.*

### 2.1 Ensure Adequate Pay and Benefits
*Among the actions that federal agencies should take are establishing and enforcing a minimum standard for pay and benefits, revising postdoctoral programs, and rethinking the apprenticeship system for graduate students.*

The funding structures that permeate academia create conditions of unequal opportunity. Many of our survey respondents reported that the academic environment is unforgiving to scientists dealing with health issues, those from low-income backgrounds, and anyone with family caregiving responsibilities. These conditions are found to have an increased negative impact on women and URMs in astronomy. A study using findings from the Longitudinal Survey of Astronomy Graduate Students (LSAGS) found that URM astronomers were more likely than non-URM astronomers to find work outside astronomy due to hostile workplace environments and inadequate compensation[6]. Compared to men, women's careers in STEM are disproportionately impacted by caregiving responsibilities. A recent study using data from the NSF's Scientists and Engineers Statistical Data System shows that 43% of women and 23% of men in STEM leave full time employment after having their first child[13]. The CSWA believes that federal funding agencies can take several actions to improve pay and benefits structures for all scientists, which will allow a more diverse group of young astronomers to remain in the field.

A recent survey of postdoctoral researchers shows they may be paid anywhere from $23,660 (the U.S. minimum wage) to over $100,000 per year, and that male researchers are paid approximately $1,700 more on average than female researchers[14]. According to the NSF's Survey of Earned Doctorates, the average postdoctoral researcher in the physical sciences receives $46,000 per year[2]. The wide range of compensation packages and low average salary (compared to what other highly educated STEM workers may earn) reflects the sentiments of many of our survey respondents, who felt it was impossible to accept such low pay with little to no benefits, especially if they had caregiving responsibilities. Federal agencies should solicit proposals for studies that will investigate the compensation of postdoctoral researchers and graduate students, and use the results to determine minimally acceptable compensation packages for each. Several of our survey respondents said that compensation needs to include health and mental health benefits that will fully support scientists throughout the stresses of finishing their degree and/or establishing themselves in a field. Agencies should update their minimum acceptable compensation standard regularly as they update their proposal and award guidelines.

Federal agencies should require that funding proposals include assurance that all team members will be compensated for their work in a way that meets these minimum standards. Additionally, agencies should revise their own postdoctoral programs' support packages to ensure they provide full and nonrestrictive healthcare and other benefits, which are not usually provided given the structure of these programs. For example, the NASA Postdoctoral Program requires postdoctoral students to have health insurance. However, NASA covers only insurance purchased through the Universities Space Research Association (USRA), and even then, the fellow must still pay approximately 15% of the insurance premium[15].

Finally, funding agencies should solicit proposals for studies that will investigate the effects of universities' funding structures on early-career scientists. The apprenticeship system for graduate students and postdoctoral students concentrates power in the hands of one advisor, creating little flexibility within a high-pressure environment. These studies should provide insight into how universities can change their funding mechanisms to better meet the needs of scientists and create healthier, more supportive work environments. If the study results are conclusive, we recommend that the agencies implement policies to respond to the study results. Alternatives to the traditional apprenticeship structure may include funding students and postdoctoral researchers through a departmental body instead of an advisor, and creating systems where students have multiple advisors within a community of mentorship.

## 2.2 Study the Two-Body Problem

*The two-body problem is found to disproportionately influence the careers of women in science. Federal agencies should initiate further study and appropriate action.*

The two-body problem refers to the issues that arise when a couple struggles to find jobs within commuting distance of each other. It is especially salient for young academic couples, who often need to move to pursue graduate school, postdoctoral fellowships, and professorships. It is a source of stress for many, as tension arises between wanting to stay with a partner and moving away to pursue what may be seen as a better opportunity. According to the StratEGIC report on dual career couples, "women report that job opportunities for their partner or spouse are important to their own career choices, and they will actively refuse job offers if their partner cannot find satisfactory employment."[16]

Women in astronomy are disproportionately affected by the two-body problem. Figure 2, from the AIP, illustrates this disparity.

The Two-Body Problem: Astronomer Couples

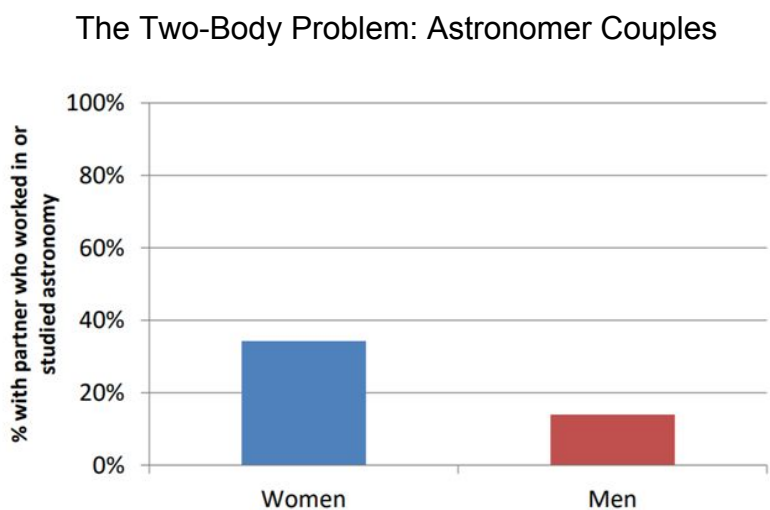

**Figure 2.** Of the participants in the Longitudinal Study of Astronomy Graduate Students, over 30% of women astronomers had a partner who is also an astronomer, whereas this is true for fewer than 20% of men surveyed.

Ivie, White, and Chu theorize that women are more likely to leave astronomy due to higher rates of contention with the two-body problem[7]. According to the AIP, women in physics and astronomy are 204% more likely than men to relocate for a spouse, and 346% more likely than men to turn down a job for a spouse[3].

The CSWA proposes that federal funding agencies solicit proposals for studies that will investigate the impacts of the two-body problem and ways to successfully mitigate it. If the results of these studies are conclusive, we recommend that the agencies implement policies to respond to the study results.

Many of our respondents expressed relief that the two-body problem was included in our survey, as they felt that much of the previous discussion around the issue did not provide any meaningful outcomes for them. More often, they were directed to simply choose between career and family, or seek a new partner. Reputable data on the impact of the two-body problem and new solutions and frameworks for viewing the issue will relieve some of the stigma surrounding it and give partnered academics windows into what a path that combines work and family priorities looks like.

2.3 Require Mentoring Plans in Proposals
*Federal agencies should require the projects they fund to include thorough mentoring plans that enable the professional development of emerging scientists.*

Federal agencies should require the inclusion of mentoring plans in proposals for all projects. The NSF already requires mentoring plans for postdoctoral scholars, which we commend and seek to expand. Receiving support from mentors is essential at every career stage, especially for women and URMs, who so often do not receive the same social support as the majority group in the workplace. We recommend that federal agencies require proposals to include mentoring plans for the undergraduate students, graduate students, postdoctoral scholars, and junior professors who are involved with a project. Mentoring plans should be treated as equally important to technical components, such as the data management plan. Standards for these plans should be most strict for long-term projects and projects with large teams, which have the potential to create many beneficial mentoring relationships.

Mentoring plans should include each senior leader's teaching and mentoring responsibilities and the process for assigning mentors. Mentoring frameworks should establish official mentoring relationships with explicit terms and meeting frequencies for mentors and mentees. Mentors should actively discuss careers with their mentees and assist them as needed with any job search. Strong mentoring plans may identify mentoring workshops and trainings that the PI and other senior leaders will attend.

### 3. Accelerating Inclusion
*To accelerate inclusion in science, federal agencies should create terms and conditions that require PIs to maintain inclusive work environments. Agencies should also continue to fund important programs such as NSF-INCLUDES and NSF-ADVANCE.*

<u>3.1 Guarantee Physical Accessibility and Inclusiveness</u>
*Funding agencies should establish a term and condition requiring PIs to detail their plans to create accessible and inclusive work environments.*

Funding agencies should ensure that all research they fund is conducted in accessible spaces by teams with inclusive practices in place by creating accessibility standards and mandating that these standards are addressed as a prerequisite for receiving funding. A proposal should be seen as sufficient when the PI has provided a plan to meet as many standards as possible within their institution's constraints, considering that not all institutions have equal resources.

These standards should be updated as proposal guidelines are updated, and agencies should coordinate their standards to the fullest extent possible. A strong accessibility plan will have provisions to accommodate people from minority gender groups, mothers, and caregivers. For example, these standards should require the PI to indicate where team members can access all-gender restrooms and lactation rooms. Teams should facilitate teleworking by employing videoconferencing software, a practice rated effective by 87% of our survey respondents.

Compliance with the Americans with Disabilities Act (ADA), is already a requirement for universities; however, agencies should nudge institutions towards exceeding ADA requirements. Simply meeting ADA requirements is often insufficient to fully accommodate all disabilities, and agencies must push institutions to go beyond simply checking off the boxes the ADA provides. For example, agencies can ask PIs to demonstrate proficiency with the assistive technologies available to them through their university and/or community, and ensure their materials use dyslexia-friendly sans-serif fonts. More strategies for increasing accessibility in academic astronomy can be found within the Nashville Recommendations, the result of the 2015 Inclusive Astronomy meeting[17]. We also encourage agencies to look to the work of AAS's WGAD for more information on advancing accessibility[18].

<u>3.2 Continue to Fund Programs and Research that Promote Inclusion</u>
*Funding agencies should continue to fund and expand their programs that create opportunities for women and minorities to succeed.*

Funding agencies should continue to fund programs that help students and young scientists pursue professional development and related opportunities. The CSWA endorses the continued funding of the NSF-INCLUDES program and supports its purpose of working to develop a national network to enhance leadership in STEM. We point to the Inclusive Graduate Education Network (IGEN) initiative as an example of a beneficial NSF-INCLUDES project. IGEN is funded under NSF-INCLUDES and will help build evidence-based knowledge about recruitment practices, with the goal of expanding the pool of candidates recruited for graduate education[19,20].

We endorse the continuing work of the NSF-ADVANCE program, which has funded successful initiatives in areas such as work-family arrangements, tenure and promotion practices, and internal networks of education, communication, and resources[21]. Funding agencies should promote and expand funding opportunities for researchers to conduct research on equal opportunity and diversity in STEM in creative ways.

The CSWA calls on the NSF to continue its work to provide scholarships and opportunities for young URM scientists through its funding programs and to collaborate with minority-serving institutions to provide opportunities for their students to enter astronomy and other fields.

### 4. Concluding Remarks
*Federal agencies should take action now to chart a path towards better support and equal opportunity for scientists.*

We invite the federal agencies, as the largest sources of funding for U.S. astronomy, to help lead our collective charge towards better diversity/inclusion outcomes by Astro2030 through the specific actionable recommendations we have outlined here. Enabling women and minorities to fully participate in science is crucial, as the U.S. population will be over 50% minorities by 2044. We call on funding agencies to take action to ensure adequate pay and benefits for graduate students and postdoctoral scholars, which will enable a more diverse pool of students to pursue graduate education. We also urge funding agencies to require inclusive practices and mentoring plans, which will help create diverse, productive, and healthy workplace environments. Continuing to investigate and mitigate the causes of structural barriers to achievement is more vital now than ever. Equity and inclusion, unlike the funding of projects and missions, is not a zero sum game: improving the diversity of astronomy as a field will lead to high-quality, impactful science. We look forward to the scientific achievements of a workforce that fully utilizes the talents of all its members.